\documentclass[aps,prl,reprint,superscriptaddress,showpacs,floatfix,longbibliography]{revtex4-1}

\usepackage{graphicx}
\usepackage{amssymb,amsmath}
\usepackage{dcolumn}
\usepackage{bm}
\usepackage[mathlines]{lineno}
\usepackage{color}
\usepackage{soul}
\usepackage{multirow}
\usepackage{amssymb,amsmath,amsthm}
\usepackage[colorlinks,linkcolor=blue,anchorcolor=blue,citecolor=blue,urlcolor=blue]{hyperref}
 \newcommand{\modR}[1] {{#1}}

\hypersetup{urlcolor=blue, citecolor=blue, linkcolor=blue,hypertexnames=false}
\begin{document}

\title{Liquid Flow through Defective Layered Membranes:\\
       A Phenomenological Description}


\author{Alexander Quandt}
\affiliation{Mandelstam Institute for Theoretical Physics
            and School of Physics,
            University of the Witwatersrand,
            2050 Johannesburg, South Africa}

\author{Andrii Kyrylchuk}
\affiliation{Institute of Organic Chemistry,
            National Academy of Sciences of Ukraine,
            Murmanska Str.~5, 02660 Kyiv, Ukraine}
\affiliation{Physics and Astronomy Department,
            Michigan State University,
            East Lansing, Michigan 48824, USA}

\author{Gotthard Seifert}
\affiliation{Theoretical Chemistry,
            Technische Universit\"{a}t Dresden,
            01062 Dresden, Germany}

\author{David Tom\'{a}nek}
\email
            {tomanek@msu.edu}%
\affiliation{Mandelstam Institute for Theoretical Physics
            and School of Physics,
            University of the Witwatersrand,
            2050 Johannesburg, South Africa}
\affiliation{Physics and Astronomy Department,
             Michigan State University,
             East Lansing, Michigan 48824, USA}

\date{\today} 

\begin{abstract}
We present a realistic phenomenological description of liquid
transport through defective, layered membranes. We derive general
expressions based on conventional models of laminar flow and
extend the formalism to accommodate slip flow. We consider
different types of defects including in-layer vacancies that
provide an activation-free tortuous path through the membrane. Of
the many factors that affect flow, the most important is the
radius of in-layer vacancy defects, which enters in the fourth
power in expressions for the flux density. We apply our formalism
to water transport through defective multilayer graphene oxide
membranes and find the flow to remain in the laminar regime. Our
results show that observed high water permeability in this system
can be explained quantitatively by a sufficient density of
in-layer pores that shorten the effective diffusion path.
\end{abstract}



\maketitle




\section{Introduction}

Porous layered structures, in particular graphite oxide
(GO)~\cite{Boehm1961}, have been claimed for a long time to bear
special advantage for the purification of liquids, in particular
the desalination of water. These initial claims have regained
recent interest due to advances in the synthesis and
characterization of such layered materials. This is particularly
true for ordered hydrophilic multilayer GO
(MLGO)~\cite{Ventrice2009}, a new term coined for layered GO,
which emerges as a very promising material for the separation of
hydrated ions from water molecules. Since liquid flow could not be
observed with sub-nanometer resolution so far, our understanding
to date has relied almost exclusively on atomistic simulations.
Only limited information is available about the microscopic
structure and the flow mechanism of water molecules in-between
layers of graphite~\cite{{Cicero2008},{DT274}} and
GO~\cite{{Korobov16},{Gogoi18},{DT274}}. %
Molecular dynamics (MD) simulations with parameterized force
fields~\cite{{Lerf06},{Wei2014},{Yang16},{Jiao2017}} and neutron
diffractions studies~\cite{Pieper06} have provided valuable
insight into the system, but significant controversies still
remain regarding the microscopic mechanism of water transport
through GO membranes. One of these involves a very high
permeability of water in MLGO~\cite{{Nair2012},{Joshi2014}}, which
could be caused by a sufficient density of atomic-scale defects,
or alternatively by the presence of `graphene capillaries' that
would cause an `ultrafast water
flow'~\cite{{Huang2013},{Nair2012}} through the membrane.


To get better insight into the matter, we have developed a
phenomenological model that addresses all relevant aspects of
liquid flow through a defective, layered membrane. Our approach is
not system specific and considers different defect geometries.
Using water flow through MLGO as an important example, we show how
the microscopic structure of the membrane affects its hydraulic
conductivity. With a sufficient density of in-layer vacancy
defects, the effective diffusion path of water molecules through
the membrane is relatively short and the resulting water
permeability is high.


\section{Formal treatment of liquid flow through defective membranes}



\subsection{Phenomenological description of constrained
            laminar flow}

The flux density $q$ of a fluid passing through a porous membrane
is given by the volume per unit area and time and has the
dimension of velocity.
For laminar flow, Darcy's law finds this quantity to be
proportional to the gradient $\nabla{p}$ %
of the applied pressure $p$, as
\begin{equation}
q=-\frac{k}{\eta}\nabla p \;. %
\label{eq:1}
\end{equation}
Here $k$ describes the permeability of the membrane,
a somewhat awkward quantity to model~\cite{Hunt2014}, and $\eta$
is the dynamic viscosity of the fluid.

\begin{figure}[t]
\centering
\includegraphics[width=1.0\columnwidth]{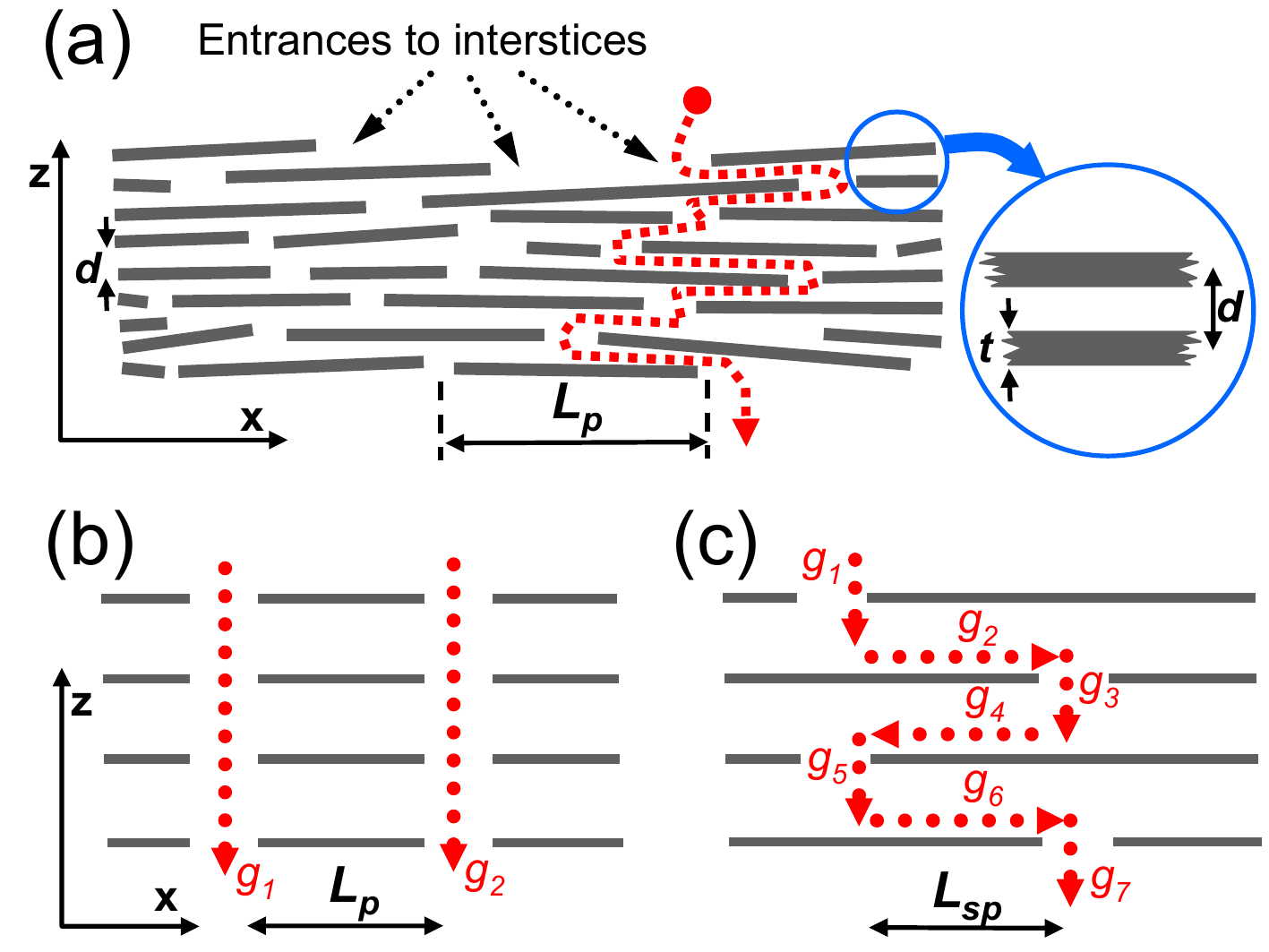}
\caption{Mechanism of liquid flow through a defective layered
membrane. %
(a) Schematic route of a molecule, proposed by
Boehm~\cite{Boehm1961} for the representative case of H$_2$O
passage through GO. The entrance to interstices occurs at the edge
of a finite flake and proceeds through slit pores in-between
layers of thickness $t$, separated by the average distance $d$.
$L_p$ is the average in-layer pore separation, with
$L_p\rightarrow\infty$ in a defect-free membrane. %
In-layer pores may be connected %
(b) in parallel and form channels that are separated by
    $L_p$ and normal to the membrane, or %
(c) in series, forming tortuous channels in a staggered %
    arrangement with lateral offset $L_{sp}$.
}%
\label{fig1}
\end{figure}

In analogy to Eq.~(\ref{eq:1}) we can define the hydraulic
conductivity $g$ of a single membrane of finite thickness $T$
using the general relation
\begin{equation}
q = g {\Delta}p \;, %
\label{eq:2}
\end{equation}
where ${\Delta}p$ is the pressure difference between the two sides
of the membrane. Next we consider a complex layered membrane
consisting of a network of pores forming a percolation path across
the membrane, as seen in Fig.~\ref{fig1}(a), which shows
schematically the flow of water through GO. The standard
approach~\cite{Hunt2014} to model the permeability of this complex
network of pores in a membrane is using a set of individual
hydraulic conductivities $g_n$ for laminar pipe flow, where the
pipe-type pores are connected either in series or in parallel, as
shown in Figs.~\ref{fig1}(b) and \ref{fig1}(c). We note that
conductivities $g_n$ are different for in-layer pores and for
``slit'' pores in-between layers.

A parallel arrangement of pores, shown in Fig.~\ref{fig1}(b),
results in an effective hydraulic conductivity $g_{par}$, whereas
an arrangement of pores in series, shown in Fig.~\ref{fig1}(c),
yields the effective hydraulic conductivity $g_{ser}$. These
effective hydraulic conductivities are given by
\begin{equation}
g_{par} = \sum_n g_n %
\label{eq:3}
\end{equation}
and
\begin{equation}
g_{ser} = \left[ \sum_n g_{n}^{-1} \right]^{-1} \;. %
\label{eq:4}
\end{equation}
For a particular system, these conductivities may be estimated
using percolation models, scaling laws and basic statistical
assumptions~\cite{Hunt2014}.

The defective membrane as a whole can be considered as a parallel
arrangement of pores that may vary in shape and density. Each pore
with its individual shape then contributes $Q_p$ to the total flux
of the liquid through the membrane.
The value $Q_p(\vec{u})$ can be uniquely characterized by the
parameter vector $\vec{u}$ for each pore, where each individual
component $u_i$ describes a quantity related to $Q_p$. Such
components may describe the spatial distribution of $Q_p$ values
along the $x-$ or $y-$direction, or the distribution of specific
factors characterizing the shape of a pore. Integration over the
entire $\vec{u}-$space should correctly account for all $Q_p$
values found in the membrane with the correct probability
distribution.

We will define the areal density of pores by $\rho_p=N_p/A$,
where $N_p$ is the number of pores across the unit area $A$. In
analogy to $Q_p$, an inhomogeneous areal density of pores can be
defined by $\rho_p(\vec{v})$, where a particular distribution of
relevant factors are described by vector components $v_j$. Also in
this case, integration over the entire $\vec{v}-$space should
correctly cover all $\rho_p$ values found in the membrane with the
correct probability distribution. This provides us with a general
expression for the flux density $q$ through a nonuniform defective
membrane
\begin{equation}
q = \frac{\int f(\vec{u}) Q_p(\vec{u}) g(\vec{v}) \rho_p(\vec{v})
               d\vec{u} d\vec{v}} %
         {\int f(\vec{u}) g(\vec{v}) d\vec{u} d\vec{v}}
  =
\modR{
  \left<Q_p{\cdot}\rho_p\right>
  }
  \;, %
\label{eq:5}
\end{equation}
where $f(\vec{u})$ and $g(\vec{v})$ are distribution functions and
the denominator provides normalization. %
\modR{ %
The purely descriptive integral expression in Eq.~(\ref{eq:5})
does not make any assumptions about defects in terms of their
density and type. It is the most general way to express the flux
density across a membrane. Among others, it provides an adequate
description of a membrane with a high density of narrow pores in
one part and a low density of wide pores in a different part. The
application of Eq.~(\ref{eq:5}) to specific types of pores is
addressed in the following.
}%

In a membrane containing uniformly distributed identical pores,
$Q_p$ and $\rho_p$ are constant, and the vector spaces
$\{\vec{u}\}$ and $\{\vec{v}\}$ collapse to single points. This
simplifies Eq.~(\ref{eq:5}) to
\begin{equation}
q %
= Q_p{\cdot}\rho_p \;. %
\label{eq:6}
\end{equation}

A standard porous membrane can be represented by a stack of
infinite sheets with an even distribution of pores. The pores may
be laterally offset by $L_{sp}$ in the staggered configuration
shown in Fig.~\ref{fig1}(c), causing tortuous flow. The tortuosity
$\tau$ can be vaguely defined by the ratio between the total
length $L_{tot}$ of a molecular trajectory and the distance
between its ends. Assuming a uniform pore distribution within a
membrane containing $N_l$ layers separated by the inter-layer
distance $d$, we find
\begin{equation}
\tau = \frac{L_{tot}}{N_l d} \;. %
\label{eq:7}
\end{equation}
We may further assume that the pores form a uniform triangular
lattice in each layer. The lateral pore separation $L_{p}$,
specified in Fig.~\ref{fig1}(b), can then be estimated using
\begin{equation}
L_{p} = \left( {\frac{2}{{3}^{1/2}\rho_p}} \right)^{1/2} \;. %
\label{eq:8}
\end{equation}
%
Further assuming an AB stacking of the triangular pore lattices in
the layered system, the lateral offset of pores in adjacent
layers, specified in Fig.~\ref{fig1}(c), is $L_{sp}=L_p/\sqrt{3}$.
With the total length of the shortest molecular path through the
membrane given by $L_{tot}=N_l(L_{sp}+d)$, the tortuosity is given
by
\begin{equation}
\tau = 1 + \frac{L_p}{\sqrt{3}d} \;. %
\label{eq:9}
\end{equation}
Our considerations so far have been general, ignoring specific
boundary conditions and the specific shape of the pores. Most
descriptions of conventional viscous flow use the no-slip theorem,
which means that the fluid adjacent to the walls of a constraint
will have the same velocity as the wall.

\subsection{Tubular pores}

There is one important pore shape, where the permeability is well
understood, namely a cylindrical pipe of finite macroscopic length
$L_{tot}$ and radius $R$. As discussed in more detail in Appendix
A, laminar flow through this pipe, subject to the no-slip theorem,
is described well by the Hagen-Poiseuille law~\cite{Hunt2014}
\begin{equation}
Q = -\frac{\pi R^4}{8\eta} \frac{\Delta p}{L_{tot}} \;. %
\label{eq:10}
\end{equation}
Here $Q$ describes the total flux, %
given by the total fluid volume passing through a cylindrical
membrane per time, driven by the applied pressure difference
${\Delta p}$. Dividing both sides of Eq.~(\ref{eq:10}) by the
cross-section of the pipe, we obtain the flux density
\begin{equation}
q = -\frac{R^2}{8\eta} {\nabla}p \;. %
\label{eq:11}
\end{equation}
Comparing this expression to Eq.~(\ref{eq:1}), the permeability
$k_t$ of a membrane containing a single tubular pore of radius $R$
is
\begin{equation}
k_t = \frac{1}{8} R^2 \;. %
\label{eq:12}
\end{equation}
As we will show below, the tubular pore or modification thereof
can also be used to describe fluid flow through nanoporous
membranes. In particular, tortuous flow through the membrane may
be represented by the flow through a multiply bent cylindrical
pipe of total length $L_{tot}$ in Eq.~(\ref{eq:10}).

\subsection{Slit pores}

A model alternative to a defective membrane is a defect-free,
layered membrane of finite lateral extent $L_{mem}$. Laminar flow
along a serpentine path through a membrane consisting of $N_l$
parallel layers separated by $d$ can also be described by the
Hagen-Poiseuille law. The flux density in this case amounts
to~\cite{Han2013}
\begin{equation}
q_{sp} {\approx} \frac{d^4}{12L_{mem}^2 \eta}
                 \frac{{\Delta}p}{N_l d} \;. %
\label{eq:13}
\end{equation}

\subsection{Beyond the no-slip theorem}

The validity of the no-slip theorem has been questioned in
nano-porous materials~\cite{{Han2013},{Holt2006}}. During slip
flow, molecules of the contained fluid may bounce along the walls
of the steric constraint, with the distance between subsequent
impacts being called the slip length. Interestingly, consideration
of slip flow does not require a major revision of the theoretical
framework presented so far~\cite{Holt2006}.

We have summarized our basic considerations regarding slip flow in
Appendix B. In the special case of slip flow in a liquid flowing
through a cylindrical pipe, the expression for the flux $Q$ in
Eq.~(\ref{eq:10}) needs to be modified to
\begin{equation}
Q = -\frac{\pi}{8\eta}\left[R^4 + L_s R^3\right]
     \frac{\Delta p}{L_{tot}} \;.
\label{eq:14}
\end{equation}
The first term in Eq.~(\ref{eq:14}) is the flux of a viscous fluid
through a pipe of radius $R$ and length $L_{tot}$ described in
Eq.~(\ref{eq:10}). The second term describes an additional flux
component due to slippage, which is characterized by the slip
length $L_s$.



\section{Characterization of defective model membranes}

\begin{figure}[t]
\centering
\includegraphics[width=1\columnwidth]{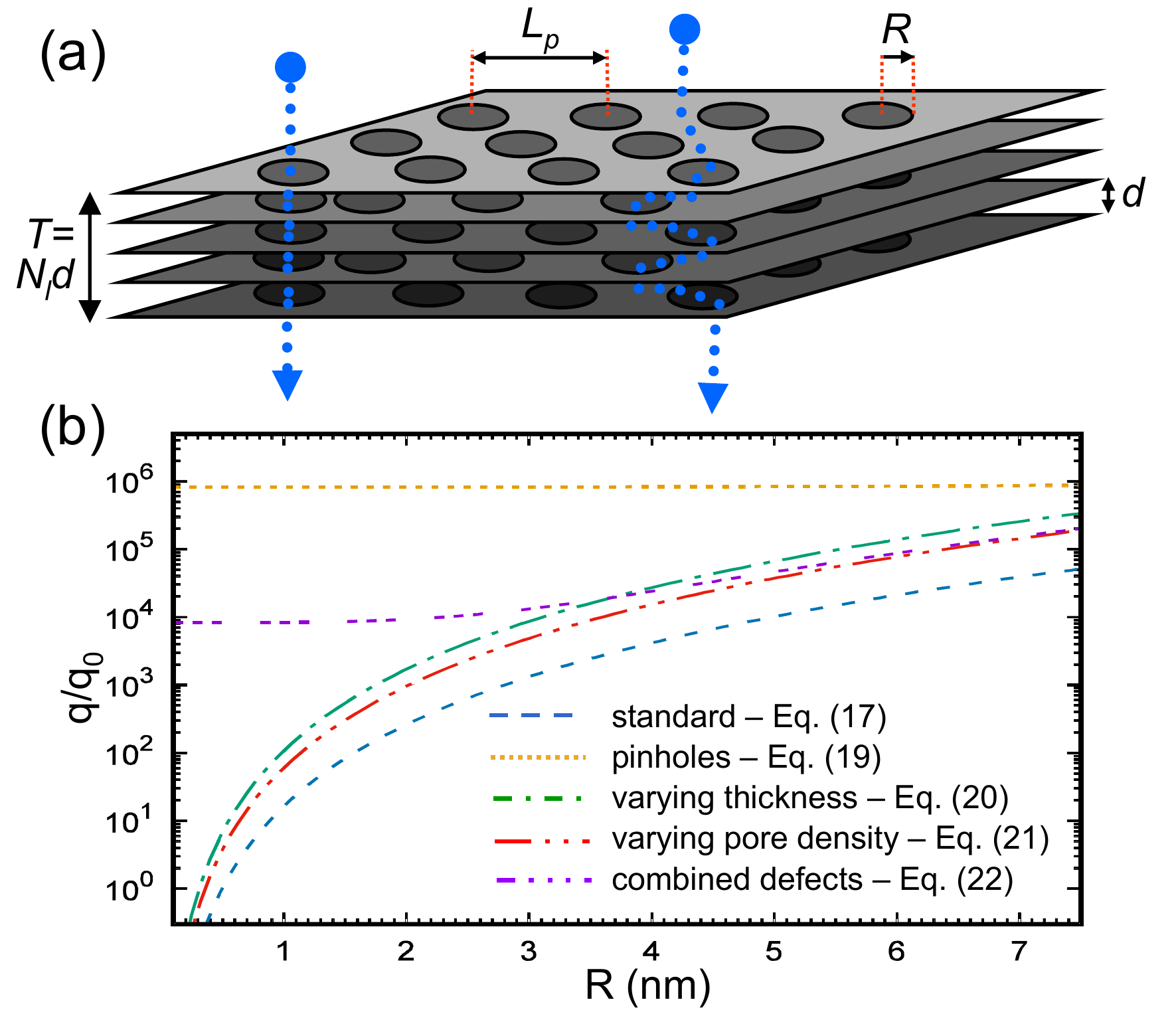}
\caption{Liquid flow through defective membranes. %
(a) Perspective view of a membrane containing $N_l$ layers,
separated by the distance $d$. Each layer contains round pores of
radius $R$ and average separation $L_p$. Straight flow through a
`pin hole', shown on the left, results from a straight arrangement
of pores. Tortuous flow shown on the right results
from a staggered arrangement of pores. %
(b) Liquid flow through defective layered membranes as a function
of pore radius $R$. Specifics of each membrane are discussed in
the text. Relative flux densities $q/q_0$ are presented in units
of the flux density $q_0$ across a reference membrane containing
pores with $R=R_0=0.5$~nm. All results are for liquid water with
the viscosity ${\eta}=10^{-3}$~Pa$\cdot$s.
}%
\label{fig2}
\end{figure}

Among the typical defects in a membrane, which we consider here,
are pores within individual layers that allow a liquid to flow
through a membrane as depicted in Figs.~\ref{fig1}(b),
\ref{fig1}(c), and \ref{fig2}(a). Among these, `pinhole defects',
shown in Fig.~\ref{fig1}(b) and Fig.~\ref{fig2}(a), provide the
shortest path for liquid flow. We consider pores with a circular
cross-section that may display a uniform or nonuniform
distribution of their radii $R$, tortuosity $\tau$, areal density
$\rho_p$, and total liquid path length $L_{tot}$. A layered
membrane, depicted in Fig.~\ref{fig2}(a), will typically contain a
combination of such defect variations that will affect the flux
density. Knowing the distribution of each of these quantities, the
flux density $q$ can be estimated using the integral
Eq.~(\ref{eq:5}). With the flux being proportional to $R^4$ in
Eq.~(\ref{eq:10}), the pore radius plays the most critical role in
the permeability of a membrane. Therefore, we will present
numerical flux densities in defective model membranes in
Fig.~\ref{fig2}(b) as a function of the pore radius. Since --
depending on the defect densities and shapes -- flux densities
change by many orders of magnitude, they are presented on a
logarithmic scale in Fig.~\ref{fig2}(b).

\subsection{Reference membrane}

What we call a `reference membrane' in Fig.~\ref{fig2} is a model
structure closely related to ${\approx}1$~$\mu$m thick GO
membranes studied by transmission electron microscopy (TEM). The
reference membrane consists of layers containing round pores of
radius $R_0=0.5$~nm, arranged in a uniform triangular lattice with
the areal pore density
$\rho_p=\rho_0=0.25{\times}10^{15}$~m$^{-2}$. The membrane
consists of $N_l=1000$ layers separated by $d=1.2$~nm, a value
typical of GO~\cite{Klechikov2015}. Using Eq.~(\ref{eq:8}) we
obtain the value $L_p=68.0$~nm for the lateral inter-pore
separation. Assuming an AB stacking of pore lattices in the
layered system and using Eq.~(\ref{eq:9}), we obtain
${\tau}_{0}=33.7$ for the tortuosity of the shortest molecular
path through the membrane. Combining Eqs.~(\ref{eq:6}) and
(\ref{eq:10}), the flux density of the reference membrane is
\begin{equation}
q_0 = -\frac{\pi R_0^4}{8\eta}
                \frac{\Delta p}{{\tau}_{0}{N_l}{\cdot}d} \rho_p \;. %
\label{eq:15}
\end{equation}
Using ${\eta}=10^{-3}$~Pa$\cdot$s for the viscosity of water and
${\Delta}p=1$~bar for a typical pressure difference between the
two sides of the membrane, we find the reference flux density
$q_0=-1.5{\times}10^{-11}$~m/s.


\subsection{Standard membrane}

What we call a `standard membrane' in Fig.~\ref{fig2} is identical
to the reference membrane with the exception that the pore radius
$R$ may vary. The flux density of the standard membrane is
then
\begin{equation}
q_{st}(R) = -\frac{\pi R^4}{8\eta}
             \frac{\Delta p}{{\tau}_{0}{N_l}{\cdot}d} \rho_p \;. %
\label{eq:16}
\end{equation}
Comparing Eq.~(\ref{eq:16}) to Eq.~(\ref{eq:15}), we find a simple
relationship between the flux densities of the standard and the
reference membrane
\begin{equation}
\frac{q_{st}(R)}{q_0} = \left( \frac{R}{R_0} \right)^4 \;. %
\label{eq:17}
\end{equation}
This relative flux density is shown in Fig.~\ref{fig2}(b) for pore
radii up to $R=7.5$~nm, the estimated average value for pores in
GO~\cite{Gaidukevic2018}.


\subsection{Membrane with pin holes}

What we call a `membrane with pin holes' in Fig.~\ref{fig2} is a
standard membrane with additional defects called `pin holes',
which cross the membrane in a straight line characterized by
${\tau}_{ph}=1$. Our description follows experimental data for
swelled GO membranes~\cite{Talyzin2014}, %
\modR{to be discussed later, }%
which contain up to micrometer-wide pin hole defects. We consider
round pin holes with a constant radius $R_{ph}=2.5$~$\mu$m and a
constant density $\rho_{ph}=10^4$~m$^{-2}$. The flux density
${\Delta}q_{ph}$ through a membrane with the parameters of the
standard membrane that is free of any defects except pinholes can
be obtained using Eq.~(\ref{eq:16}) and amounts to, in units of
$q_0$,
\begin{equation}
\frac{{\Delta}q_{ph}}{q_0} = \left( \frac{R_{ph}}{R_0} \right)^4 %
                        \frac{\rho_{ph}}{\rho_0} \ {\tau}_{0} \;.
\label{eq:18}
\end{equation}
The flux density through a standard membrane containing additional
pinholes is then given by
\begin{equation}
\frac{q_{ph}(R)}{q_0} = \frac{q_{st}(R)}{q_0} %
                      + \frac{{\Delta}q_{ph}}{q_0} \;. %
\label{eq:19}
\end{equation}
The constant quantity ${\Delta}q_{ph}/q_0=0.82{\times}10^6$ is so
large that it dominates over $q_{st}(R)/q_0$ in the entire radius
range considered in Fig.~\ref{fig2}(b).


\subsection{Membrane with varying thickness}

The number of layers may not be the same across defective layered
membrane. Let us consider a standard membrane with $N_l=1000$
layers. A fraction $\alpha$ of this membrane is much thinner,
containing only $N_{l,f}$ layers. Then, according to
Eq.~(\ref{eq:10}), the relative flux density will be
\begin{equation}
\frac{q_{vt}(R)}{q_0} =
\frac{q_{st}(R)}{q_0} %
\left[ 1 + {\alpha} \left( \frac{N_l}{N_{l,f}}-1 \right) \right] \;. %
\label{eq:20}
\end{equation}
For $\alpha=30\%$ and $N_{l,f}=100$ we obtain a $3.7$ times higher
flux density than for a standard membrane. This result is shown in
Fig.~\ref{fig2}(b).


\subsection{Membrane with varying pore density}

Also the areal density $\rho_p$ of uniform pores may vary across
the membrane. Let us consider a standard membrane with
$\rho_p=\rho_0=0.25{\times}10^{15}$~m$^{-2}$. In a fraction
$\beta$ of this membrane, the areal pore density $\rho_{p,f}$ is
different from the rest. In that case, according to
Eq.~(\ref{eq:16}), we get
\begin{equation}
\frac{q_{vd}(R)}{q_0} =
\frac{q_{st}(R)}{q_0} %
\left[ 1 + {\beta} \left( \frac{\rho_{p,f}}{\rho_p}-1 \right) \right] \;. %
\label{eq:21}
\end{equation}
Assuming a local pore density increase by a factor of $15$ in a
fraction $\beta=40\%$ of the membrane, we find a flux density
increase by a factor of $6.6$ over the standard membrane. This
result is shown in Fig.~\ref{fig2}(b).


\subsection{Membrane with a combination of defects}

A typical membrane will contain a combination of defects. We
consider a standard membrane with a fraction $\alpha$ containing a
different number of layers $N_{l,f}$, a fraction $\beta$
containing a different pore density $\rho_{p,f}$, and a fraction
$\gamma$ containing pin holes, with
${\alpha}+{\beta}+{\gamma}{\le}1$. The relative flux density
through such a membrane is given by
\begin{eqnarray}
\frac{q_{cd}(R)}{q_0} &=& %
  {\alpha} \frac{q_{vt}(R)}{q_0} %
+ {\beta}  \frac{q_{vd}(R)}{q_0} %
+ {\gamma} \frac{q_{ph}(R)}{q_0} \nonumber \\ %
&+& (1-{\alpha}-{\beta}-{\gamma}) \frac{q_{st}(R)}{q_0} \;,%
\label{eq:22}
\end{eqnarray}
where %
$q_{vt}(R)/q_0$ is given by Eq.~(\ref{eq:20}), %
$q_{vd}(R)/q_0$ by Eq.~(\ref{eq:21}), and %
$q_{ph}(R)/q_0$ by Eq.~(\ref{eq:19}). %
Results for such a membrane containing defect combination with
${\alpha}=30\%$, %
${\beta}=40\%$ and %
${\gamma}=1\%$ %
are presented in Fig.~\ref{fig2}(b).


\subsection{Membrane with a Gaussian distribution of pore radii}

\begin{figure}[h]
    \centering
    \includegraphics[width=0.9\columnwidth]{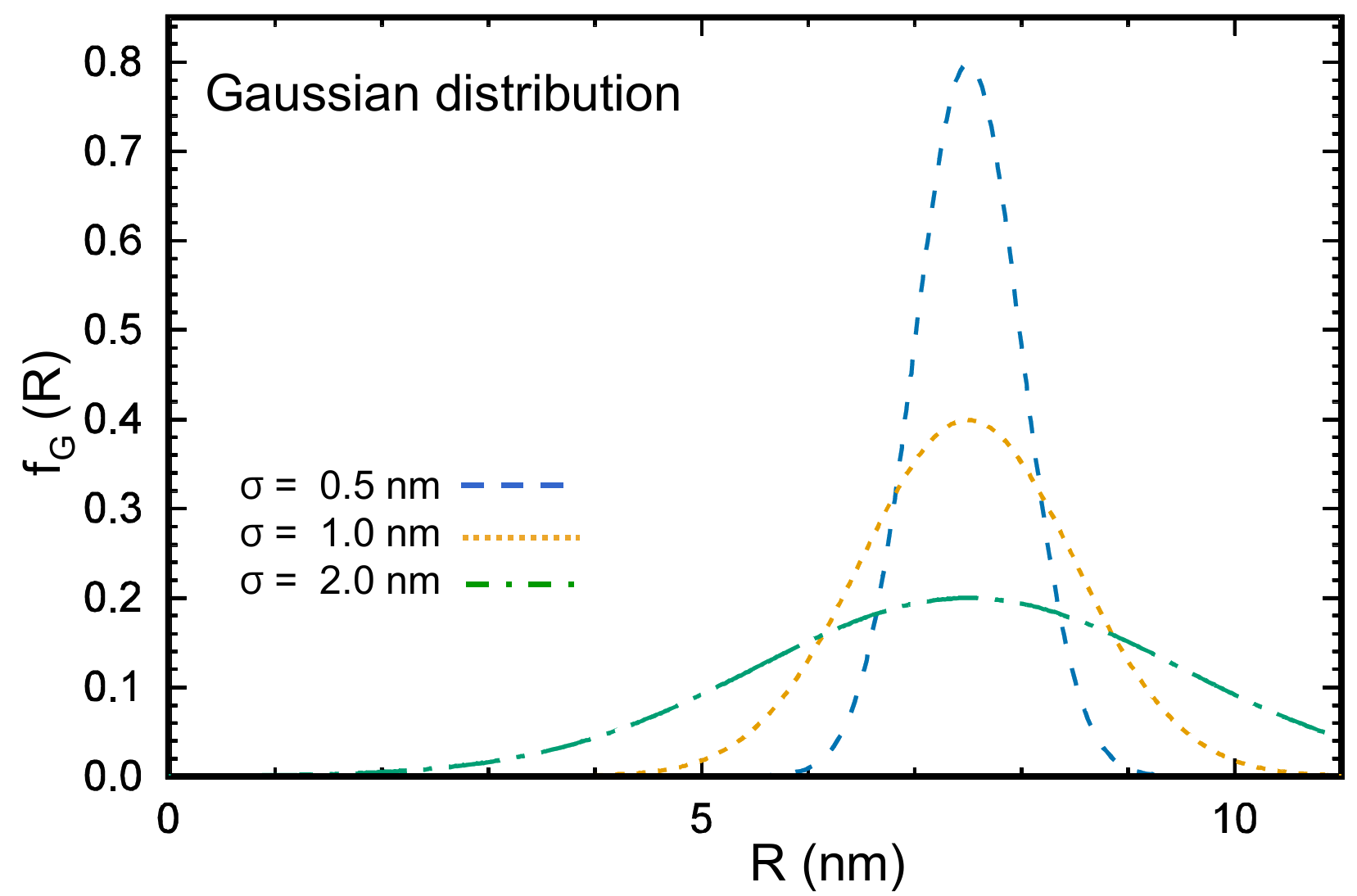}
    \caption{Fraction of pores with specific radii represented by
    a Gaussian distribution defined in Eq.~(\ref{eq:23}).
    Distributions are presented for the average radius
    $\left<{R}\right>=7.5$~nm and different values of $\sigma$.
    }%
    \label{fig3}
\end{figure}

For a constant value of the pore radius $R$, the relative flux
density through a standard membrane is given by Eq.~(\ref{eq:16}).
We have to go back to Eq.~(\ref{eq:5}) if the pore radii are not
constant, but rather represented by a probability distribution
function $f(R)$. %
\modR{ %
In the following, we consider pores with a narrow distribution of
radii around the mean value. %
}%
A Gaussian or normal distribution of radii is described by
\begin{equation}
f_G(R) = \frac{1}{\sigma\sqrt{2\pi}} %
       e^{-\frac{1}{2}\left( \frac{R-<R>}{\sigma} \right)^2} \;,
\label{eq:23}
\end{equation}
where $\left<{R}\right>$ is the average radius, $\sigma$ is the
standard deviation, and the full width at half maximum (FWHM) is
$2.355\sigma$. The pre-exponential factor provides for
normalization. Results for $\left<{R}\right>=7.5$~nm and different
values of $\sigma$ are presented in Fig.~\ref{fig3}.

The extension of Eq.~(\ref{eq:15}) for the expectation value of
the relative flux density in case of a Gaussian radius
distribution is
\begin{equation}
\left< \frac{q_{st}}{q_0} \right> %
    = \frac{1}{R_0^4}
      \int_0^\infty f_G(R)R^4 dR \;,
\label{eq:24}
\end{equation}
where $f_G(R)$ is the expression (\ref{eq:23}) for a Gaussian
distribution of radii. We evaluated the quantity $<q_{st}/q_0>$
for the Gaussian distribution with $\left<{R}\right>=7.5$~nm.
Results for three different values of $\sigma$ used in
Fig.~\ref{fig3} are presented in Table~\ref{table1}.

As expected, results of Eq.~(\ref{eq:24}) converge to those of
Eq.~(\ref{eq:17}) for ${\sigma}{\rightarrow}0$.

\begin{table}[b]
\centering
\caption{%
Relative flux density through a standard membrane containing
different distributions of round pores. Results presented are
obtained using Eq.~(\ref{eq:24}) for the Gaussian distribution and
Eq.~(\ref{eq:26}) for the Cauchy distribution of radii.
}%
    \begin{tabular}{lcccr} %
    \hline \hline
     \multicolumn{2}{c}{Gaussian distribution}
   & \hspace*{1cm}
   & \multicolumn{2}{c}{Cauchy distribution} \\
     ${\sigma}$~(nm)
   & $\left<{q_{st}/q_0}\right>$
   &
   & ${\nu}$~(nm)
   & $\left<{q_{st}/q_0}\right>$ \\
   \hline
\modR{ $0.5$ } & \modR{ $5.2{\times}10^4$} & & \modR{ $0.5$} & \modR{ $8.7{\times}10^8$ } \\
\modR{ $1.0$ } & \modR{ $5.6{\times}10^4$} & & \modR{ $1.0$} & \modR{ $1.7{\times}10^9$ } \\
\modR{ $2.0$ } & \modR{ $7.3{\times}10^4$} & & \modR{ $0.1$} & \modR{ $3.4{\times}10^9$ } \\
   \hline \hline %
    \end{tabular}
    \label{table1}
\end{table}

\subsection{Membrane with a Cauchy distribution of pore radii}

\begin{figure}[h]
\centering
\includegraphics[width=0.9\columnwidth]{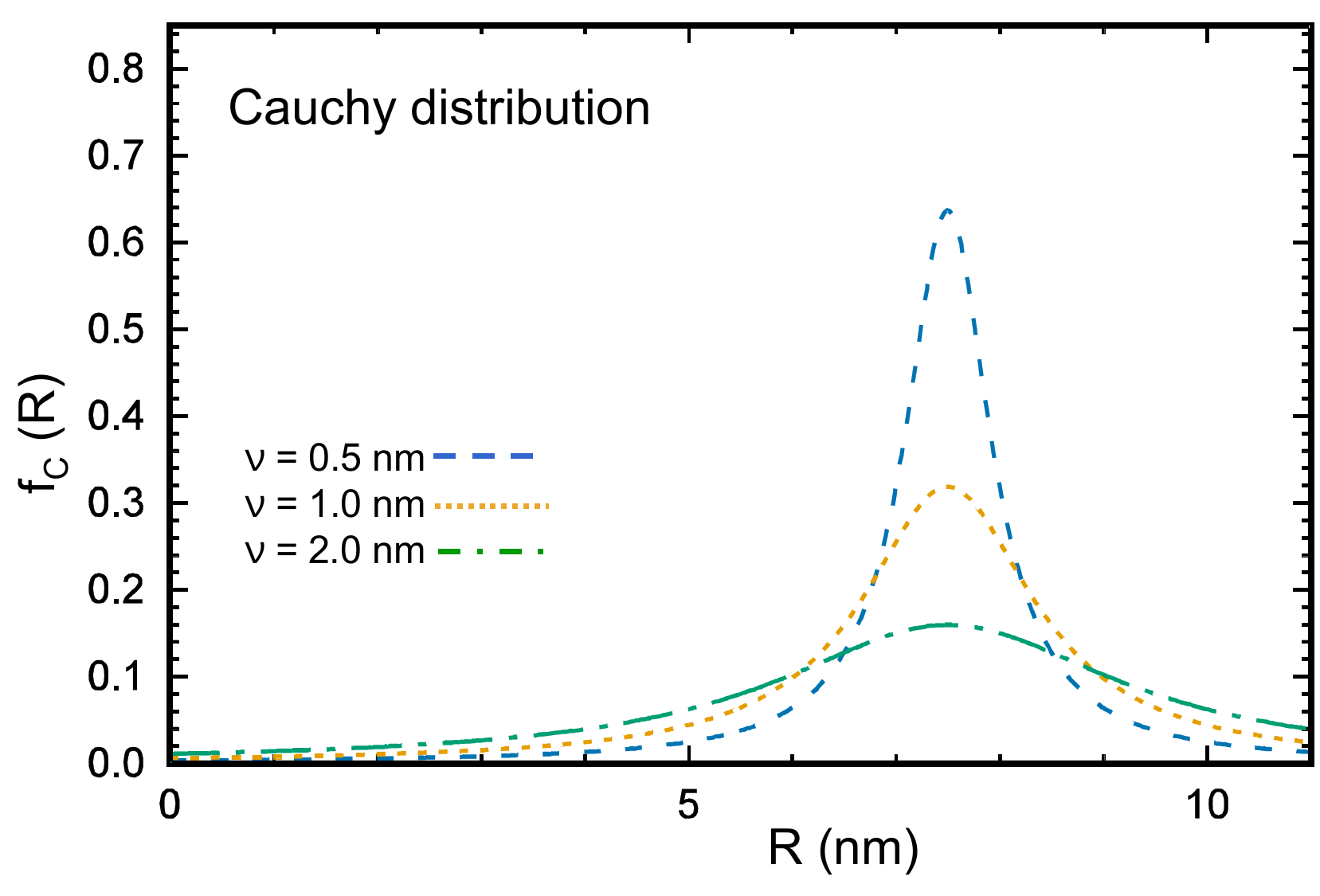}
\caption{Fraction of pores with specific radii represented by
    the Cauchy distribution defined in Eq.~(\ref{eq:25}).
    Distributions are presented for $\left<{R}\right>=7.5$~nm
    and different values of $\nu$.
}%
\label{fig4}
\end{figure}

Instead of the Gaussian distribution of radii $f_G(R)$ defined in
Eq.~(\ref{eq:15}), we may consider the Cauchy distribution
\begin{equation}
f_C(R) = \frac{1}{\pi\nu %
         \left[1 + \left(\frac{R-<R>}{\nu} \right)^2\right]} \;, %
\label{eq:25}
\end{equation}
where $\left<{R}\right>$ is the most probable radius and $2\nu$ is
the FWHM. This distribution carries more weight in the tails
around $\left<{R}\right>$ than the Gaussian distribution. $f_C(R)$
is shown in Fig.~\ref{fig4} for $\left<{R}\right>=7.5$~nm and
different values of $\nu$.

The logical way to obtain the relative flux density through a
standard membrane with a Cauchy radius distribution is to replace
$f_G(R)$ by $f_C(R)$ in Eq.~(\ref{eq:24}), leading to
\begin{equation}
\left< \frac{q_{st}}{q_0} \right> %
    = \frac{1}{R_0^4}
      \int_{0}^{R_{max}} f_C(R)R^4 dR \;.
\label{eq:26}
\end{equation}
The important difference between Eqs.~(\ref{eq:24}) and
(\ref{eq:26}) is the upper limit of the integration range. Even
though $f_C(R)$ is normalized, its fourth moment diverges for
$R_{max}{\rightarrow}\infty$. We consider the finite value
$R_{max}=10^3$~nm large enough in comparison to $\left<{R}\right>$
to provide us with reasonable estimates of
$\left<{q_{st}(R)/q_0}\right>$ for the Cauchy distribution of
radii.

Results for three different values of $\nu$ used in
Fig.~\ref{fig4} are presented in Table~\ref{table1}.
We note that flux densities are many orders of magnitude higher
with the Cauchy distribution of radii than with the Gaussian
distribution for the same value of $\left<{R}\right>$.



\section{Relevant example: GO membrane}

\subsection{Morphology of a defective GO membrane}

GO and MLGO membranes are considered excellent candidates for the
filtration of liquids. MLGO is a highly aligned structure of
layered GO, which consists of graphene layers that are chemically
functionalized by epoxy-O and OH groups~\cite{Boehm1961}. Unlike
hydrophobic graphite and graphene, GO is hydrophilic and requires
water for stabilization. The two commonly used approaches to
synthesize GO are the Hummer~\cite{Hummers1958} and the
Brodie~\cite{Brodie1859} process. Both techniques yield an ordered
material containing a substantial fraction of in-layer pores. Due
to the intrinsic defective nature of GO, its designation as
``graphene oxide'' is, strictly speaking, a misnomer.

GO has been introduced already in the 1960's as a suitable
material to form membranes that are particularly useful for water
desalination based on reverse osmosis~\cite{Boehm1961}. This
material has been shown to find applications in the filtration and
purification of not only water~\cite{Akhavan10}, but also other
substances, which have to cross many pores in many layers.

A realistic structure of GO membranes, displayed schematically in
Fig.~\ref{fig1}(a), is very different from an ideal stack of 2D
layers and varies from sample to sample. Membranes should be
considered a 3D system with a strong heterogeneity in the size,
shape and orientation of GO flakes that contain %
\modR{ %
a variety of in-layer defects~\cite{{Talyzin2014},{Pacile2011},%
{Erickson2010},{Si2008},{Loh2010}}
}%
including vacancies.

\subsection{Water transport through defective GO:
            Merits of atomistic and phenomenological
            descriptions}

Proper description of water transport through GO membranes is
highly desirable, also to clarify the origin of the reported high
flux density of water in these
membranes~\cite{{Huang2013},{Nair2012}}. The conventional
description~\cite{Boehm1961} of water transport through a porous
multilayered GO membrane assumes a rather direct way of water
molecules along the ``slit'' pores in-between GO flakes or through
possible holes in the GO structure. The more recent alternative
interpretation~\cite{Nair2012,Joshi2014} of the high permeability
of GO postulates a relatively long path, including transport along
the inter-layer slit pores of the GO structure. According to the
authors, a high water flow rate can then only be explained by
introducing a new fundamental mechanism called superpermeability.
Key to this mechanism is so-called ultrafast flow of water through
graphene nanochannels~\cite{{Huang2013},{Nair2012}}, which have
been postulated to be present in GO membranes. These claims have
been disputed to a large degree~\cite{Talyzin2018} based on
microscopic structure studies of GO. Meanwhile, a number of
experimental and theoretical studies have been published to better
understand this rather exotic transport
mechanism~\cite{Huang2013,Cho2017,Han2013}.

Due to the complexity of the problem, atomistic molecular dynamics
(MD) simulations have typically used empirical force fields that
reproduce the structure of liquid water well, but usually failed
to describe the interaction of water molecules with GO layers with
the same accuracy. Yet it is exactly the interplay between
water-water and water-GO interactions that is key to the
understanding of water transport through pores in GO layers and
even more along the slit pores. Moreover, as mentioned before, the
realistic structure of defective GO is very complex. In that case,
the predictive power of MD simulations addressing specific
geometries is limited.

Under these circumstances, the phenomenological formalism in this
study offers a viable alternative to describe water transport
through GO membranes from a statistical viewpoint. This approach
is based on well-established evidence that water properties
including viscosity and diffusion constants are not changed during
laminar flow through a nanopore. Also the interaction between
water molecules and GO layers does not change in a GO
nanopore~\cite{DT274}. This approach offers a bias-free view of
water flow through many types of tubular and slit pores that
likely are present in the GO structure~\cite{Talyzin2018} and that
have been discussed above.


\subsection{Scenarios of water transport through defective GO}

\subsubsection{Tubular pores in GO}

An insightful description of flow through nano-porous materials
has been provided~\cite{Holt2006} for a membrane that is
impermeable except for pores consisting of aligned carbon nanotube
(CNT) arrays, which alone allow for the passage of fluid
molecules. This corresponds to the case of hydraulic
conductivities in parallel described by Eq.~(\ref{eq:3}) and shown
in Fig.~\ref{fig1}(b). For a typical pressure difference
${\Delta}p{\approx}1$~bar between the two sides of the
$T=1.2$~$\mu$m thick GO reference membrane, we obtain
$|q|{\approx}|q_0|=1.5{\times}10^{-11}$~m/s for the flux density
when using the Hagen-Poiseuille expression (\ref{eq:15}) for
laminar flow through round tubes in absence of slip flow.


\subsubsection{Slit pores in GO}

In case of lateral flow following a labyrinthine path through slit
pores in-between defect-free GO layers, the flux density is
described by Eq.~(\ref{eq:13}). Using $L_{mem}{\approx}1~\mu$m for
the lateral extent of the layers in an otherwise defect-free
reference membrane, we obtain
$q_{sp}{\approx}1.4{\times}10^{-9}$m/s.


Our estimated results for the flux density through tubular pores
are six orders of magnitude and those for slit pores four orders
of magnitude~\cite{Han2013} lower than the measured flux densities
of water in GO~\cite{{Huang2013},{Nair2012}}. The discrepancy has
been attributed to slip flow~\cite{{Huang2013},{Nair2012}}, with
estimated slip lengths of the order of $L_s{\approx}0.1-1$~$\mu$m.
A viable alternative explanation for the difference, discussed in
this manuscript, is based on the presence of in-layer defects that
have been overlooked.


\section*{Discussion}

The purpose of our study was to introduce a general heuristic
framework to describe the flow of fluids through a membrane
consisting of a defective layered material. Such membranes find an
important application in separating unwanted substances from
liquids. We focused on nanoporous GO membranes that bear promise
for the desalination of water. Clearly, the same framework will
also apply in many other cases.

The motivation for our study was the observation of high water
flux rates through GO membranes and the controversy about their
interpretation. We were especially disturbed by the discrepancy of
several orders of magnitude between observed flux densities and
our calculated values based on the assumption of laminar flow and
the adequacy of the Hagen-Poiseuille law.

This is the reason why we used Occam's razor approach to revisit
all assumptions used in our modelling study. As discussed in
Appendix A, there is no doubt that fluid flow in GO membranes is
laminar and its description by the Hagen-Poiseuille law is
adequate. The only remaining point in question is the presence and
characteristics of defects.

Defects in the membrane structure may have been caused by severe
damage that somehow went undetected. Pores in the structure may be
more numerous or have different shapes than estimated. Since not
all pores may be easily recognized in electron microscopy
observations, pore densities obtained in such studies are likely
underestimated. Additional uncertainty about realistic pore
densities stems from the fact that microscopic pieces of the
membrane used in microscopy may not be representative of the
entire system.

Of all the factors that affect flow through defects most, we found
that the pore radius $R$ is by far the most important, since flux
depends on $R^4$. It is quite likely that pore sizes have been
underestimated, since pores are usually characterized by an
electron microscope in dry samples. In the natural aqueous
environment, water penetrates the membrane, swells it and widens
the pores. We note that increasing the pore radius $R$ by a mere
factor of %
\modR{ %
$2$ %
}%
should increase the flux density by a factor of %
\modR{ %
$16$ %
}%
according to the Hagen-Poiseuille law. Furthermore, swelling
could increase the internal pressure within the membrane, but
should not change the effect of the applied pressure ${\Delta}p$
driving the flow.

Based on present experimental evidence, we believe that GO
membranes are defective to a higher degree than previously
admitted. Thus, we conclude that the significant discrepancy
between observed high flux densities~\cite{{Huang2013},{Nair2012}}
and numerical values in our study can be explained simply using
Hagen-Poiseuille law for laminar flow through different types of
defects that either were not detected or their cross-section was
underestimated.

Other interpretations of the apparent discrepancy between Theory
and Experiment have been discussed extensively in the
literature~\cite{{Cho2017},{Han2013},{Talyzin2018}}, so that we
may only offer a brief summary in the following.


Alternative interpretations of the observed ``superfast'' flow of
``superfluid'' water between GO
sheets~\cite{Joshi2014,Nair2012,Yang2017b} are based on specific
postulates regarding the microscopic structure of the membrane.

{\em Graphene capillaries}, consisting of large interconnected
unoxidized areas of GO sheets~\cite{Nair2012,Joshi2014,
Yang2017b,Abraham2017}, have been postulated to increase the flow
of water through GO membranes. Even though the inter-layer spacing
$d$ in graphite is only ${\approx}3.3$~{\AA} and thus much smaller
than in GO, $d$ has been postulated to increase significantly to
${\approx}7-7.5$~{\AA} in graphene capillaries located in oxidized
areas. A large network of graphene capillaries was postulated to
maintain slip-flow of water and thus provide record-breaking
permeability.

So far, however, there has been no independent evidence of such
capillaries or even presence of pristine graphene in
GO~\cite{Talyzin2018} based, among others, on high-resolution TEM
imaging~\cite{Erickson2010} and Raman spectra~\cite{Feicht2018}.
Presence of graphene capillaries should favor He permeation in
water-free membranes rather than block it, as observed
\cite{Nair2012}. If graphene capillaries were present and
important, permeation rates should depend on the flake sizes in
laminates rather than being constant~\cite{Saraswat2018}. Also,
presence of hydrophobic graphene areas within GO should increase
entry/exit barriers, which were specifically
considered~\cite{Belin2016} and discussed~\cite{Walther2013} in
the context of water transport through hydrophobic CNT-based
membranes. It appears that a good strategy to maximize flow would
be to reduce or eliminate activation barriers at the entrance and
exit of water molecules. Such favorable conditions exist at the
edges of hydrophilic GO membranes that are free of pristine
graphene.

{\em Ideal microscopic membrane} consisting of stacks of
defect-free micrometer-sized GO layers at a ${\approx}0.7$~nm
inter-layer distance was postulated as another important factor
favoring ``superfast'' water flow~\cite{Nair2012}. However, it is
well established that defects including pores and cracks are
formed during sonication of GO, the preferred method for obtaining
GO nanosheets~\cite{Liscio2017}. This fact suggests that the
postulate of a defect-free GO membrane is unrealistic.


Our own interpretation of the observed high permeation rate, based
on what is known about GO membranes, their structure and
preparation, is much simpler: the high flow rate may be attributed
to nanostructured defects that provide many short transport
channels in parallel.


\section*{Summary and Conclusions}

We have developed a realistic phenomenological description of
liquid transport through defective, layered membranes. To do so,
we have revisited the established formalism describing laminar
flow and extended it to accommodate slip flow, a theoretical
concept that would provide large corrections to laminar flow
though defects. We applied the general expressions we derived to
layered membranes containing a variety of defects that are not
system-specific. As a representative example, we chose layered GO,
an important hydrophilic membrane system used for purification of
water. Using realistic data for GO membranes, we estimated the
force densities acting on water and found them to be insufficient
to push the system out of the laminar regime, making the concept
of slip flow redundant.

We found that in-layer vacancies, which prevail in defective GO,
open a short, but tortuous path for activation-free transport of
molecules through the membrane. Of the many factors that affect
the flow, the most important is the radius of vacancy defects,
which enters in the fourth power in the expression for the flux
density based on the Hagen-Poiseuille law. In thin membranes with
a sufficient density of such in-layer defects, the effective
diffusion path of water molecules is relatively short and the
water permeability increases significantly. We believe that a
discrepancy of four orders of magnitude between observed water
flow in GO membranes and calculated flux densities based on the
Hagen-Poiseuille law are well explained by the presence of
in-layer defects. In view of the fact that pores open up and swell
in a membrane exposed to a liquid, electron microscopy
observations on dry samples would typically underestimate the
density and effective size of defects. This explains the
significant increase in flow density without the need to invoke
the presence of ``graphene capillaries'' that would enable an
``ultrafast flow rate'' of water.


\section{Appendix}
\setcounter{equation}{0}
\renewcommand{\theequation}{A\arabic{equation}}

\subsection{A. Laminar flow through tubular pores}

In the following we provide additional background
information~\cite{White2003} used in the derivation of the
Hagen-Poiseuille law describing laminar flux through a round pipe.
We start our discussion with the most fundamental equation of
fluid flow~\cite{deGennes2004}
\begin{equation}
   \vec{f}=\rho\frac{d\vec{v}}{dt} \;.
\label{eq:A1}
\end{equation}
Here, $\vec{f}$ is the net force per unit volume acting on the
fluid of mass density $\rho$ and velocity $\vec{v}$. Of course,
Eq.~(\ref{eq:A1}) is Newton's second law restated for a fluid.
Since the problem of pipe flow is quasi one-dimensional, we will
drop the vector notation in the following.

The right-hand side of Eq.~(\ref{eq:A1}) represents the inertial
force per volume $f_i$. The net force density $f$ %
consists of a driving force density $f_d$ and a viscous drag force
density $f_v$. For a fluid flowing through a pipe with average
velocity $\left<{v}\right>$, the three forces can be estimated
by~\cite{deGennes2004}
\begin{equation}
f_i  = \rho \frac{d{v}}{d{t}} %
                  {\approx} \rho\frac{\left<{v}\right>^2}{L_{tot}}  %
\label{eq:A2}
\end{equation}
and
\begin{equation}
f   = f_d + f_v  {\approx} - \frac{{\Delta}p}{L_{tot}}
                             + \eta \frac{\left<{v}\right>}{R^2}
                  \;.
\label{eq:A3}
\end{equation}
The meaning of all the other terms is the same as in
Eq.~(\ref{eq:10}).

We note that for non-vanishing $f_i$, the physics resulting from
Eqs.~(\ref{eq:A1}) and (\ref{eq:A2}) will be nonlinear due to the
quadratic dependence of $f_i$ on the average fluid velocity
$\left<{v}\right>$. As we will show in the following,
$|f_i|<<|f_d|, |f_v|$ under the conditions discussed here, so that
the inertial force density may be neglected.

Let us first estimate the magnitude of the relevant force
densities under the simplifying assumption that the average fluid
velocity $\left<{v}\right>$ and the typical flux density $q$ are
of similar orders of magnitude. For water flow through a standard
membrane with $R=1$~nm, described by Eq.~(\ref{eq:16}), we obtain
$\left<{v}\right>{\approx}|q_{st}|=2.4{\times}10^{-10}$~m/s.
According to Eq.~(\ref{eq:A2}), the inertial force density
$f_i={\rho}\left<{v}\right>^2/L_{tot}=1.4{\times}10^{-12}$~N/m$^3$
if we use $\rho=10^3$~kg/m$^3$ for the density of water and
$L_{tot}=40.4$~$\mu$m for its tortuous path through the membrane.
The viscous force density, defined in Eq.~(\ref{eq:A3}), is
$|f_v|={\eta}\left<{v}\right>/R^2=2.4{\times}10^5$~N/m$^3$.
Considering the pressure difference ${\Delta}p=1$~bar between the
two sides of the membrane, assumed for the reference membrane, and
using Eq.~(\ref{eq:A3}), we obtain
$|f_d|={\Delta}p/L_{tot}=2.4{\times}10^9$~N/m$^3$. Since $|f_i|$
is $17-21$ orders of magnitude smaller than $|f_d|$ and $|f_v|$,
it can be safely neglected. This justifies our assumption that
that flow though membranes we consider is well within the laminar
regime. An independent confirmation of laminar flow obeying the
Hagen-Poiseuille law at even higher pressure differences
${\Delta}p=6$~kbar has been provided by atomistic MD simulations
of water flow through model nanopores~\cite{Goldsmith09}.

Under steady-state conditions, $f$ must vanish and we conclude
that $f_i{\approx}f=0$. Then, we may adopt some of the expressions
presented above for the specific case of laminar flow through a
round nano-pipe. Assuming radial symmetry, the flow velocity $v$
is described by the differential equation
\begin{equation}
    \frac{1}{\eta}\frac{\Delta p}{L} = \frac{d^2 v}{dr^2} +
                                       \frac{1}{r}\frac{dv}{dr}
 = \frac{1}{r}\frac{d}{dr} \left(r\frac{dv}{dr}\right) \;.
    \label{eq:A4}
\end{equation}
A general solution has the form
\begin{equation}
    v(r) = \frac{1}{4\eta} r^2 \frac{\Delta p}{L} +
           C \cdot \ln(r) + D \;.
    \label{eq:A5}
\end{equation}
Values of the parameters $C$ and $D$ are determined assuming axial
symmetry and suitable boundary conditions at the walls. For
conventional flow, the assumption of axial flow leads to
${dv}/{dr}=0$ at $r=0$ representing the center of the pipe. The
no-slip condition at $r=R$ at the wall implies that $v(R)=0$.

Once the parameters $C$ and $D$ are set, the flux $Q$ may finally
be obtained via a straight-forward integration, which leads to the
Hagen-Poiseuille-law
\begin{equation}
    Q = \int_{0}^{R} 2 {\pi} r v(r) dr = -\frac{\pi R^4}{8\eta}
        \frac{{\Delta}p}{L_{tot}}
    \label{eq:A6}
\end{equation}
that is identical to Eq.~(\ref{eq:10}).

\subsection{B: Slip flow through tubular pores}

In the case of slip flow, we have to abandon the no-slip condition
$v(R)=0$. Instead, we impose a new type of boundary condition,
which describes a scenario, where the fluid molecules are bouncing
along the walls of the pipe with the velocity
\begin{equation}
    v(R) = L_s \cdot \frac{dv}{dr}(R) \;.
    \label{eq:A7}
\end{equation}
Integrating over the radius with the modified boundary condition
in Eq.~(\ref{eq:A7}) leads to the expression
\begin{equation}
Q = -\frac{\pi}{8\eta}\left[R^4 + L_s R^3\right]
     \frac{\Delta p}{L_{tot}} \;.
\label{eq:A8}
\end{equation}
that is identical to Eq.~(\ref{eq:14}). Note that the slip length
$L_s$, which occurs in this phenomenological boundary condition,
is a parameter that may be freely adjusted to explain any observed
deviation from the Hagen-Poiseuille law for laminar flow through a
pipe.


\section*{Author contributions}

A.Q., G.S. and D.T. conceived the idea. A.Q. and D.T. wrote the
manuscript. A.K. researched literature and performed selected
calculations. D.T. finalized the manuscript with feedback from the
other authors.


\section*{Acknowledgments}

\begin{acknowledgments}
D.T. acknowledges financial support by the NSF/AFOSR EFRI 2-DARE
grant number EFMA-1433459. A.Q. and D.T. acknowledge support by
the Materials for Energy Research Group (MERG), the DST-NRF Center
of Excellence in Strong Materials (CoE-SM) at the University of
the Witwatersrand, the Mandelstam Institute for Theoretical
Physics (MITP) and the Simons Foundation, award number 509116. We
appreciate valuable discussions with Alexandr Talyzin, Igor
Baburin and Dan Liu. Computational resources have been provided by
the Michigan State University High Performance Computing Center.
\end{acknowledgments}


%

\end{document}